\documentstyle{elsart}
\begin{document}
\begin{frontmatter}
\title{Power fluctuations in the wavelet space: large-scale
CMB non-gaussian statistics}
\author{L. Popa }
\address{Institute for Space Sciences, Bucharest-Magurele, 
R-76900, Romania }

\begin{abstract}
We analyse the large-scale coherence of the CMB anisotropy field
with non-gaussian initial conditions  using 2-point 
function of the power fluctuations
in the wavelet space.\\
Employing the multivariate Edgeworth expansion (MEE) 
we constrain the normalization for the cosmic string mass-per-unit-length
$\mu$, obtaining $G \mu /c^{2}= 1.075^{+0.455}_{-0.375} \times 10^{-6}$
at $68\%$ CL in the standard gaussian statistics. This value is consistent 
with the results obtained from the simulations of the evolution of 
the string network and by other large scale studies.
\end{abstract}


\end{frontmatter}

\section{Introduction}

The cosmic microwave background (CMB) provides one of the most
useful tools for understanding the physical processes in the early universe,
giving valuable information about the origin and the dynamics
of the primordial fluctuations that seeded the today large-scale
structures.
There are two competing families of cosmological theories 
quite distinct in this regards.\\
In the models invoking most types of inflation 
\cite{Guth81,Linde82,Liddle93}
the perturbations' generation mechanism
is linear and gaussian fluctuations are predicted.
For a given type of inhomogeneity (adiabatic or isocurvature)
the only degree of freedom in the initial conditions is the amplitude, 
and since
the evolution is linear, solutions scale linearly with the
initial amplitude.
Linearity combined with the assumed statistical homogeneity of the
universe guarantees that one may decompose the cosmological perturbations
into eigenmodes with eigenfunctions that evolve independently.
In the Fourier space, for instance, the evolution of a perturbation
in each mode $k$ is independent. 
For a  finite sample of the CMB field,  the 
observed fluctuations are the convolution of the ``true" fluctuations of the 
infinite field with the Fourier transform of a mask \cite{Feldman94}.
The 2-point function  of the  the power fluctuations 
depends only on  the selection function \cite{Feldman94,Stirling96}.\\  
The second class of theories, most topological defect models
\cite{Vilenkin85,Vilenkin94,Coulson94}
and some versions of inflation \cite{Kofman96},
predict non-gaussian statistics of the primordial perturbations.
Those models involve spontaneous symmetry breaking  as the hot
early universe cooled leading to  a non-linear perturbations'
generation mechanism.
For this type of perturbations the Fourier modes $k$
are coupled and their evolution cannot be anticipated on the basis of the
initial conditions \cite{Albrecht96,Ferreira97}.
The predicted angular power spectrum has a position dependence with a variance 
larger than the variance of the angular power spectrum for the gaussian case
at the same multipole order \cite{Magueijo95,Ferreira97}.
The defect theories are highly constrained by causality, the fluctuations
being completely uncorrelated beyond the horizon scale at all times
\cite{Albrecht96,Crittenden96}.\\
It is difficult to make difference between gaussian and non-gaussian 
fluctuations of the density field at large angular scales because of the 
tendency
of any distribution to approach gaussian statistics when averaging over
large area
(the central limit theorem), and also because of the cosmic and sampling
variances.\\
In this paper we propose the wavelet transform to analyse the 
large scale coherence of the CMB anisotropy field 
with non-gaussian initial conditions.  
The main advantage of the wavelet transform over the Fourier
transform  is its capability to investigate 
features of the CMB anisotropy maps with a 
resolution according to their scales  \cite{Chui92}.
The wavelet base functions are highly
localized in space and the information regarding the scale and the position
are stored explicitly. Also, the wavelet analysis 
can be performed with a finite sample of data 
in contrast with the Fourier
analysis where the result depends on the selection function.\\
We employ  2-point function of power fluctuations in
the wavelet space to compare  COBE-{\em DMR} 
4-year temperature anisotropy
maps to the predictions of an analytical cosmic string model
\cite{Perivolaropoulous93}.
The advantage of the  analytical 
approach  is that one  can obtain statistical fluctuations 
with non-gaussian random phases for  a given experimental configuration,
while the explicit 
dependences on the normalization of the primordial power spectrum,
string parameters and angular scales can 
be shown.\\  
We use  the non-gaussian likelihood function to constrain
the dimensionless cosmic string parameter $G \mu/c^{2}$ (where $G$ is the 
Newton's
constant, $c$ is the speed of light and
$\mu$ the the mass per unit length of
the cosmic strings) 
at $68\%$ CL in  standard gaussian statistics, by comparing zero lag 
autocorrelation function from COBE-{\em DMR} with the predictions 
of the cosmic string model. 

\newpage
\section{Monte-Carlo simulations}

According to the scaling solution for cosmic strings 
\cite{Bennet88,Allen90} there is a
fixed number $M$ of strings with the curvature radius 
of the order of the
horizon per Hubble volume at any given time, with orientations and
velocities uncorrelated over distances larger than the horizon.
The temperature anisotropy induced by long strings is given by 
Kaiser-Stebbins formula \cite{Stebbins84}:
$$ \frac {\delta T}{T}= \pm 4 \pi G \mu | \hat{k} \cdot(\gamma_{s}v_{s}
 \times \hat{e}_{s})|, $$
where: $\hat{k}$ is the direction of observation, $v_{s}$ is the velocity of
the string with the orientation $\hat{e}_{s}$,
$\gamma_{s}=(1-v_{s}/c)^{1/2}$, $G$ is the Newton's constant and 
$\mu$ is the
mass per unit length of the string.\\
The time from  the  
last scattering surface until now
is divided into $N$ Hubble time steps $t_{i}$
with $t_{i+1}=2t_{i}$ and $N=log_{2}(t_{0}/t_{ls})$. 
The apparent angular size of a Hubble volume at time $t_{i}$ is
$\theta_{H_{i}} \sim z_{i}^{-1/2} \sim t_{i}^{1/3}$ and 
$\theta_{H_{i+1}}=2^{1/3} \theta_{H_{i}}$ 
for large redshifts in the matter dominated era
assuming $\Omega_{0}=1$.\\
For an experiment measuring a square map of size 
$\theta^{0} \times \theta^{0}$  
at each Hubble time $t_{i}$ the number of 
$n_{i}$ string segments random placed over the area size 
$(\theta^{0}+\theta_{H_{i}})^{2}$ is given by \cite{Moessner94}:
\begin{equation}
n_{i}=M(\theta^{0}+\theta_{H_{i}})^2/\theta_{H_{i}}^{2}.
\end{equation}
We use  an analytical statistical string model
\cite{Perivolaropoulous93} to take into account the combined effects of
the temperature fluctuations induced by strings present 
between the redshift $z \simeq 250$ and today ($N=12$).\\
We calculate the probability distribution functions
of the temperature fluctuations
induced by cosmic strings for different values of the string scaling
solution parameter $M$ by Fourier transforming the characteristic function
of  this model
normalized to approach a  gaussian distribution with zero 
mean and $\sigma=1$
when $M \rightarrow  \infty$ \cite{Perivolaropoulous93}.\\
We generated CMB independent realizations, each of 6144 pixels, 
using the spherical harmonic representation:
\begin{equation}
\frac {\delta T}{T}(\theta, \phi)=\sum^{30}_{l=2} \sum^{l}_{m=0}
a_{lm}W_{l}Y_{lm}(\theta,\phi).
\end{equation}
where $W_{l}$ is the DMR gaussian window function with FWHM=$7^{\circ}$.\\ 
The harmonic coefficients $a_{lm}$ are random variables with zero mean
and the variances given by \cite{Bond87}:
\begin{equation}
<a^{2}_{lm}>=Q^{2}_{rms-PS} \frac{4 \pi}{5}
\frac{\Gamma[l+(n-1)/2]\Gamma[(9-n)/2]}
     {\Gamma[l+(5-n)/2]\Gamma[(3+n)/2]},
\end{equation}
where: $Q_{rms-PS}$ is the quadrupole normalized amplitude and  $n$ is the
spectral index of the primordial power spectrum.\\
The coefficients $a_{lm}$
are drawn from parent populations obtained in the
cosmic string model 
The non-gaussian amplitude distributions of $a_{lm}$ with random
phases convoluted with the spherical harmonics $Y_{lm}(\theta, \phi)$ 
are characterized by
a positive kurtosis in the resulting temperature distribution.\\
We generated 800 $n=1$ full-sky realizations for each model defined by
$Q_{rms-PS}$ and $M$.
To each realization we added a realization of the noise
determined by the instrument sensitivity and 
the number of observations per pixel
of  DMR 4-year 53GHz (A+B)/2 map,  and rejected the pixels with galactic
latitude $-19^\circ<b<20^\circ$ obtaining $2^{12}$ pixels.\\
We  generated also in the same conditions 
800 $n=1$ full sky realizations
for the same values of
$Q_{rms-PS}$ with  $a_{lm}$ drawn from a standard gaussian
distribution with zero mean and $\sigma=1$.

\section{Two point function of power fluctuations
in the wavelet space}

A wavelet \cite{Chui92} $\Psi(x) \in R$
is a function whose binary dilatation and dyadic translations generate 
an orthonormal base  so that any function
$f(x) \in R$ can be approximated up to an arbitrarily small
precision by:
\begin{equation}
f(x)= \sum_{i \in N } \sum^{2i-1}_{j=0}
                c^{i}_{j} \Psi(2^{i}x-j).
\end{equation}
The base functions $\Psi(x)$
are labeled by the scale $i$ and the position  $j$ $(i,j \in N)$. 
The wavelet coefficients $c^{i}_{j}$
are real quantities containing all the information as that contained by the
Fourier coefficients. The major difference is the number of indices, that
is, in wavelet analysis  the scale $i$
and the position $j$ in  the scale are explicitly stored.
The wavelet base functions $\Psi(x) $ satisfy the scaling equation:
\begin{equation}
\Psi^{i}_{j}(x)=2^{-i/2} \Psi(2^{-i}(j-x)).
\end{equation}
Let $V_{H_{i}}$ be the Hubble volume spanned by the
scaling function $[\Phi(2^{i}x-j)]_{j \in N}$ at the scale $i$.
From the scaling equation it follows that:
$$ ... \subset V_{H_{i-1}} 
       \subset V_{H_{i}} 
       \subset V_{H{i+1}} 
       \subset ...$$
This hierarchical structure ensures the multi-resolution capability of 
the wavelet analysis.\\  
The orthogonality condition:
\begin{equation}
\left< \Psi^{i}_{j} \Psi^{i^{'}}_{j^{'}} \right>
		=\frac{1}{2^{i}}\delta_{j,j^{'}} \delta_{i,i^{'}},
\end{equation}

yields to:
\begin{equation}
c_{j}^{i}=2^{i}\sum_{k}f(x_{k}) \Psi^{i}_{j}(x_{k}).
\end{equation}
This last equation represents our wavelet transform convention.\\
For a random gaussian field  the ensemble-averaged power 
$\hat{P}(\Delta \theta)$  
in the wavelet space at the separation angle $\Delta \theta$ is:
\begin{equation}
<c_{j}^{i}(\vec{n_{1}}) 
  c_{j^{'}}^{i^{'}}(\vec{n_{2}})>_{\Delta \theta}=
	\delta_{i,i^{'}} \delta_{j,j^{'}}\hat{P}(\Delta \theta),
\end{equation}
where $\vec{n_{1}}$ and $\vec{n_{2}}$ are two directions in 
the sky separated by $\Delta \theta$.\\
For a random non-gaussian field generated using the cosmic string 
model the intrinsic correlations 
exist only within the same scale and the equation (8) becomes: 
\begin{equation}
<c_{j}^{i}(\vec{n_{1}}) 
  c_{j^{'}}^{i^{'}}(\vec{n_{2}})>_{\Delta \theta}=
	\delta_{i,i^{'}} \hat{P}_{jj^{'}}(\Delta \theta),
\end{equation}
and $\hat{P}_{jj^{'}}(\Delta \theta)$ contains the correlated 
ensemble-averaged signal.\\
We define the 2-point function of the power fluctuations in the wavelet 
space as:
\begin{equation}
\xi_{P}(\Delta \theta)=
		\delta \hat{P}_{jj^{'}}(\Delta \theta)= 
	\hat{P}_{jj^{'}}(\vec{n_{1}}) \hat{P}_{jj^{'}}(\vec{n_{2}})-
	\hat{P}_{jj^{'}}(\Delta \theta).
\end{equation}	 
In this paper we  present the results obtained using Daubechies
wavelet base functions of order 4 as a typical compact orthonormal
base.\\
The wavelet transform  applied  on  2D map (with $2^{12}$ pixels
and $i=12$ uncorrelated scales) acts in a pyramidal algorithm,
stripping the initial map
scale by scale into its correlated components. 
The reader may refer
to the references \cite{Press92,Fujiwara96} for complete information.\\
Figure 1 presents the distributions of spots  
as  a function of threshold (expressed in number of standard deviations)
obtained for  the averaged simulated maps in real and wavelet space
for two types of simulations: $a_{lm}$ with  random gaussian phases
(panels a and b ), and  $a_{lm}$ with random non-gaussian  phases,
obtained in the cosmic string model with $M=10$ (panels c and d ).
For all simulations we take $Q_{rms-PS}=18\mu$K. 
In each case the gaussian fits are also shown.
One can see that the differences between the distributions 
obtained in gaussian and non-gaussian case 
are more evident in the wavelet space than in the real 
space.\\ 	
The maps  obtained by Monte-Carlo simulations were wavelet transformed and
then 2-point  functions 
of the power fluctuations were constructed (binned in $n=36$ bins, 
in the manner of COBE-{\em DMR}).\\
Figure 2 presents the 2-point function of the 
power fluctuations in the wavelet space 
$\xi_{P}(\Delta \theta)$ normalized at $\xi_{P}(0)$.
The solid curve represents the theoretical prediction if the underlying
distribution is gaussian and  $Q_{rms-PS}=18\mu$K.
The error bars are based on Monte-Carlo simulations in the cosmic 
string model with $M=10$ and the same normalization. 
  
\section{Non-Gaussian Likelihood analysis}

According to the standard $\chi^{2}$ method, if a specific model 
defined by the parameters $(Q_{rms-PS},M)$ is more likely to 
have occurred   higher is the probability to obtain values of the
$\chi^{2}$ larger than the measured $\chi^{2}_{0}$. 
The cumulative probability used to define the confidence 
regions on the parameters is given by the integral:
\begin{equation}
I(\chi^{2}_{0})=\int_{0}^{\chi^{2}_{0}} L(\lambda, \chi^{2}) d \chi^{2},
\end{equation}
where ${\em L}$ is the likelihood function  full specified by 
the $\chi^{2}$ value and the covariance matrix.
We define the $\chi^{2}$ as:
\begin{equation}
\chi^{2}=\sum_{i=1}^{36} \sum_{j=1}^{36}(<\xi^{i}_{P}>-{\xi^{i}_{P}}^{DMR})
\lambda^{-1}_{ij}(<\xi^{j}_{P}>-{\xi^{j}_{P}}^{DMR}).
\end{equation}
where $<\xi_{P}>$ is the ensemble-averaged value 
of the 2-point function of the power
fluctuations for Monte-Carlo realizations and $\xi_{P}^{DMR}$
is the two point function of the power fluctuations for
{\em DMR} data.\\
The covariance matrix $\lambda_{ij}$ calculated for the
the Monte-Carlo realizations is:
\begin{equation}
\lambda_{ij}= \frac {1} {N_{realiz}} \sum_{k=1}^{N_realiz} 
( \xi^{k}_{P}-< \xi_{P}> )( \xi^{k}_{P}-< \xi_{P} >).
\end{equation}
The standard Gaussian likelihood statistic $L_{g}$  associated to the
two point function of the power fluctuation $\xi_{P}( \Delta \theta)$  
is given by:
\begin{equation}
L_{g}=\frac {  \e^ {- \frac{1}{2} \chi^{2} } }
{(2 \pi)^{n/2}( det \lambda)^{1/2}}.
\end{equation}
In the case of non-gaussian random fields the confidence regions
of the parameters depend also on the higher order correlation functions.
It is shown  
that for mild non-gaussianity
the likelihood function 
can be obtained within the multivariate Edgeworth expansion (MEE) 
\cite{Kendall 1987}.
According to MEE the total likelihood function is:
\begin{equation}
{\em L}=L_{g}+L_{ng},
\end{equation}
where $L_{ng}$ is the  non-gaussian correction which embodies the 
higher order moments of the data.\\
The integral in equation (11) can be written as:
\begin{equation}
I({\chi^{2}_{0}})=F_{n}(\chi^{2}_{0})+Q_{n}(\chi^{2}_0),
\end{equation}
where $F_{n}(\chi^{2}_{0})$ is the integral of gaussian likelihood, 
that is   
the cumulative integral of the $\chi^{2}$ probability distribution  
with $n=36$ degrees of freedom, and $Q_{n}(\chi^{2}_{0})$
is a non-gaussian correction that depends on the higher order moments.\\ 
For each set of  Monte-Carlo simulations 
we estimated the cumulative probability $I(\chi^{2})$.\\
Figure 3 presents  $I(\chi^{2})$ obtained for the models with 
$Q_{rms-PS}=18\mu$K  and few values of the $M$ parameter compared with
the same probability obtained for the gaussian case (the upper curve).\\
One can see that for mild non-gaussian fields  
the confidence regions are larger than
the confidence regions in the gaussian case 
at the same $\chi^{2}$ threshold.\\
Using the gaussian likelihood given by equation (14) we found 
$Q_{rms-PS}=18.29 \pm 1.36 \mu K$ when the DMR 53GHz (A+B)/2
map was compared with the Monte-Carlo simulations obtained
for a  gaussian underlying distribution.\\
Figure 4 presents the confidence regions obtained for the parameters
$Q_{rms-PS}$ and $M$ in cosmic string  models at 
$\chi^{2}_{0}=31.1$ that corresponds to $68\%$ CL  
for a purely gaussian statistics. \\
Considering the region with  $I(\chi_{0}) \geq 0.3$
and taking the extrema of this contour 
we found that DMR data favourize the cosmic string  
models with $M$ within 10 and 12 and $Q_{rms-PS}$ within
$17.76 \mu K$ and  $20.24 \mu K$ at $68\%$ CL 
for  the standard gaussian statistics.
We estimated the dimensionless cosmic parameter $G\mu/c^{2}$ 
by comparing,  in the wavelet space, the averaged 
zero lag autocorrelation function obtained in those models 
with zero lag autocorrelation function of
DMR 4-year $53GHz$ $(A+B)/2$ map.
We found:
\begin{equation}
G \mu /c^{2}= 1.075^{+0.455}_{-0.375} \times 10^{-6} . 
\end{equation}
There are two terms that contribute to the error: a symmetrical error
($\pm 0.185$) originating from the width of the confidence region
and an asymmetric error  given by the cosmic and sampling variances.

\section{Conclusions}

In this paper we propose the wavelet analysis for  studying the large-scale
CMB correlations induced by the cosmic strings.\\ 
Using an analytical cosmic string model to generate
CMB temperature fluctuations with non-gaussian random phases
in the COBE-{\em DMR} experimental configuration, and  
2-point function of the power fluctuations in the wavelet space,
we constrain the normalization of the primordial power spectrum 
$Q_{rms-PS}$ and the parameter of the cosmic string scaling solution $M$.\\
We found that the large-scale CMB anisotropy favourizes the cosmic string 
models with large number of strings per Hubble volume, $M$=10, 11 and 12
and $Q_{rms-PS}$ from $17.76\mu$K to $20.24\mu$K at $68\%$ CL
for  the standard gaussian statistics.\\
By comparing 
zero lag autocorrelation function of COBE-{\em DMR} 53GHz (A+B)/2 map
with our predictions we obtained the dimensionless
cosmic string parameter 
$G \mu/c^{2}=1.075^{+0.455}_{-0.375} \times 10^{-6}$, value  
that is compatible with the values obtained
from the simulations of the evolution of the string network 
from $z=100$ to present \cite{Allen96},
as well as with other existing studies of the expected
large-scale CMB anisotropy 
\cite{Perivolaropoulous93}, \cite{Bennet92,Hara93}.

\ack{This work was partially supported by M.C.T. grant 3005GR.
The COBE-{\em DMR} data sets, developed by NASA Goddard Space 
Flight Center under the guidance of the COBE Science Working group,
were provided by the NSSDC.}


\begin{thebibliography}{99}

\bibitem{Guth81}
A.H. Guth, Phys. Rev. {\bf D 23} (1981) 347
\bibitem{Linde82}
A. Linde, Phys. Lett {\bf B 108} (1982) 389
\bibitem{Liddle93}
A. Liddle and D. Lyth, Phys. Rep.  {\bf 231} (1993) 1
\bibitem{Vilenkin85}
A. Vilenkin, Phys. Rep. {\bf 121} (1985) 263.
\bibitem{Vilenkin94}
A. Vilenkin and E.P.S. Shellard {\em Cosmic Strings and Other 
Topological Defects} (Cambridge University Press, 1994)
\bibitem{Coulson94}
D. Coulson, P. Ferreira, P. Graham, 
and N. Turok Nature {\bf 368} (1994) 27
\bibitem{Kofman96}
L. Kofman, A. Linde and A.A Starobinsky Phys. Rev. Lett. {\bf 76}
(1996) 1011
\bibitem{Albrecht96}
A. Albrecht, D. Coulson, P. Ferreira and J. Magueijo Phys. Rev. Lett
{\bf 76} (1996) 1413
\bibitem{Magueijo95}
J. Magueijo, A. Albrecht, D. Coulson and P. Ferreira 
Phys. Rev. Lett. {\bf 76} (1996) 2617
\bibitem{Ferreira97}
P. Ferreira and J. Magueijo Phys. Rev. {\bf D 55} (1997) 3358
\bibitem{Crittenden96}
R. Crittenden and N. Turok Phys. Rev. Lett. {\bf 75} (1995) 2642
\bibitem{Kendall 1987}
M. Kendall, A. Stuart and J.K. Ord {\em Kendall's Advanced Theory 
of Statistics } (Oxford University Press, New York, 1987)
\bibitem{Feldman94}
H.A. Feldman, N. Kaiser, J.A. Peacock Ap. J. {\bf 426} 23.
\bibitem{Stirling96}
A.J. Stirling and J.A. Peacock MNRAS {\bf 283L} (1996) 99
\bibitem{Chui92}
C.K. Chui, {\em An Introduction to Wavelets} (Academic Press, 1992)
\bibitem{Press92}
W.H. Press, B.P. Flannery, S.A. Teukolsky, W.T. Vetterling
{\em Numerical Recipes} (Cambridge University Press, 1992)
\bibitem{Bennet88}
D. Bennet and F. Bouchet, Phys. Rev. Lett. {\bf 60} (1988) 257
\bibitem{Allen90}
B. Allen and E.P.S. Shellard Phys. Rev. Lett. {\bf 64} (1990) 119 
\bibitem{Stebbins84}
N. Kaiser and A. Stebbins Nature {\bf 310} (1984) 391
\bibitem{Moessner94}
R. Moessner, L. Privolaropoulos, and R. Branderberger, 
Ap. J. {\bf 425} (1994) 365
\bibitem{Perivolaropoulous93}
L. Perivolaropoulos, Phys Lett. {\bf B298} (1993) 305
\bibitem{Bond87}
J.R. Bond and G. Efstathiou, MNRAS {\bf 226} (1987) 655
\bibitem{Fujiwara96}
Y. Fujiwara and J. Soda, Prog. Theor. Phys. {\bf 95} (1996) 1059
\bibitem{Allen96}
B. Allen, R.R. Caldwell, E.P.S. Shellard, A. Stebbins 
and S. Veeraraghavan Phys. Rev. Lett. {\bf 77} (1996) 3061
\bibitem{Bennet92}
D. Bennet, A. Stebbins, and F. Bouchet, Ap.J.Lett  {\bf 399} (1992) L5
\bibitem{Hara93}
T. Hara, P. Mahonen, and S.Miyoshi, Ap.J {\bf 414} (1993) 412
\end{thebibliography}
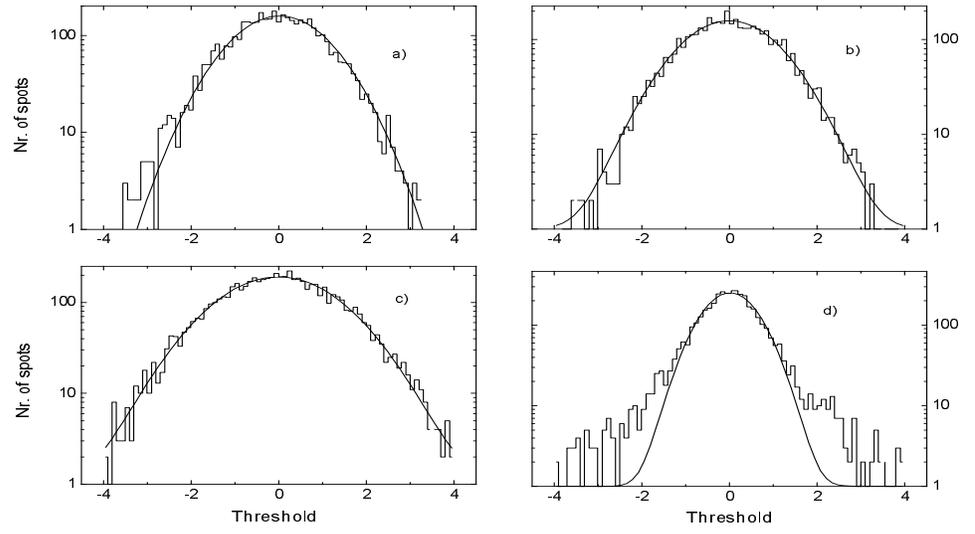
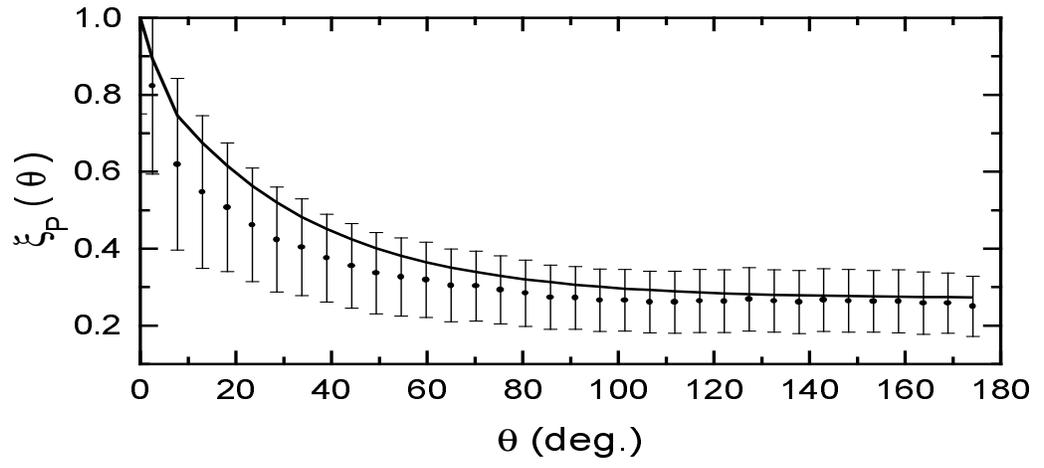
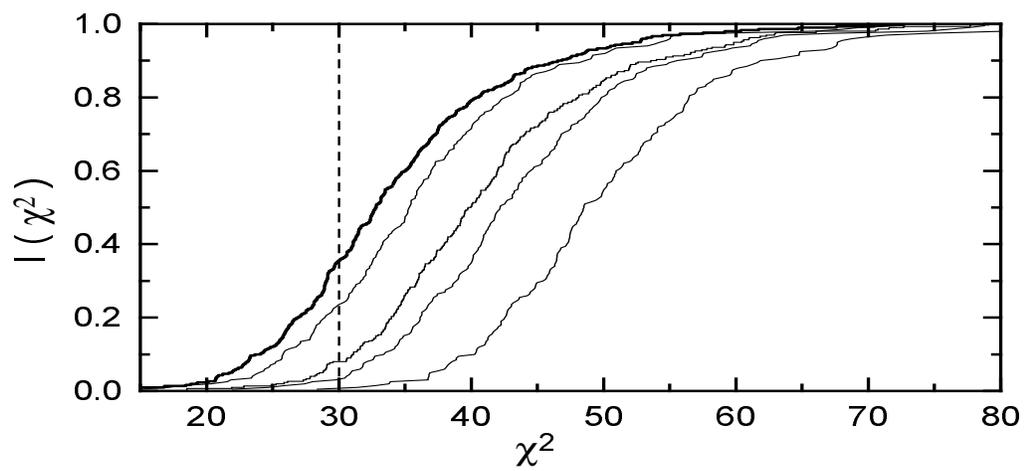
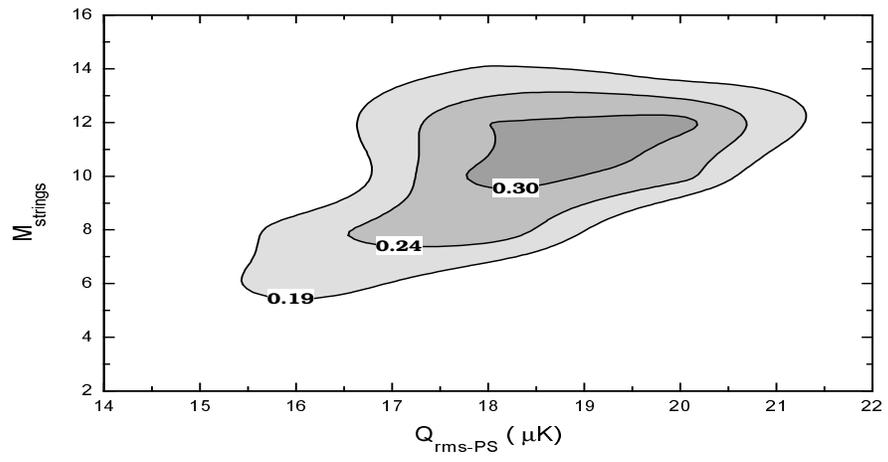
\end{document}